\documentstyle{article}

\newcommand{\q}{{\bf u}}
\newcommand{\vk}{{\bf e}_z}
\newcommand{\vi}{{\bf e}_x}
\newcommand{\x}{{\bf x}}
\newcommand{\k}{{\bf k}}
\newcommand{\cG}{{\cal G}}

\begin{document}

\title{\bf The Swift-Hohenberg equation requires non-local modifications
to model spatial pattern evolution of physical problems}
\author{A.J.~Roberts\thanks{Department of Mathematics \& Computing,
University of Southern Queensland, Toowoomba, Queensland 4350,
Australia.  E-mail: {\tt aroberts@usq.edu.au}}}
\date{9th December, 1994}
\maketitle

\begin{abstract}
 I argue that ``good'' mathematical models of spatio-temporal dynamics in
two-dimensions require non-local operators in the nonlinear terms.
Consequently, the often used Swift-Hohenberg equation requires modification
as it
is purely local. My aim here is to provoke more critical examination of the
rationale for using the Swift-Hohenberg equations as a reliable model of
the spatial pattern evolution in specific physical systems.
\end{abstract}

\section{Introduction}

Consider the spatio-temporal dynamics of systems with a very large
horizontal extent, when compared to their height. For two definite
examples, I will
refer to Rayleigh-Benard convection,
\begin{equation}
	\begin{array}{rcl}
		\frac{1}{Pr}\left(\frac{\partial\q}{\partial t}+\q\cdot\nabla\q\right)
		 & = & -\nabla p'+Ra\,\theta\vk +\nabla^2\q\,,  \\[1mm]
		\frac{\partial\theta}{\partial t}+\q\cdot\nabla\theta & = &
		w+\nabla^2\theta\,,
	\end{array}
	\label{rbeq}
\end{equation}
where $Pr$ and $Ra$ are the Prandtl and Rayleigh numbers respectively,
and also refer to a toy set of partial
differential equations,
\begin{equation}
	\begin{array}{rcl}
		\frac{\partial a}{\partial t} & = & ra-\left(1+\nabla^2\right)^2a-ab\,,
		\\[1mm]
		\frac{\partial b}{\partial t} & = &
		-\left(\alpha_0-\alpha_1\nabla^2+\alpha_2\nabla^4\right)b+a^2\,.
	\end{array}
	\label{toyeq}
\end{equation}
For values of the parameters near some critical value, a continuum of modes,
an annulus in Fourier space $|\k|\approx\mbox{const}$, become linearly
unstable.
Through physical nonlinearities these modes not only saturate, but also
interact. The interaction of the continuum of critical modes is the main
feature of interest as it determines the spatial patterns in the
horizontal. In order to explore pattern evolution, many researchers
\cite[for
example]{Brand88,Cross84,Cross86,Eckmann91,Elphick91,Greenside84,Pomeau80}
have invoked the Swift-Hohenberg \cite{Swift77} equation,
\begin{equation}
	\frac{\partial A}{\partial t}=\mu A-\left(k_0^2+\nabla^2\right)^2A
	-\gamma A^3\,,
	\label{sheq}
\end{equation}
as a useful system to study. Whereas it is indeed instructive to examine
the Swift-Hohenberg equation, I contend that the Swift-Hohenberg equation
is deficient as an accurate and reliable model of most specific physical
planform evolution problems.

Instead, I propose that it is generally necessary to incorporate non-local
nonlinearities into a low-dimensional model of planform evolution.
Specifically, I recommend an equation of the form
\begin{equation}
	\frac{\partial A}{\partial t}=\mu A-\left(k_0^2+\nabla^2\right)^2A
	-A\cG\star A^2\,,
	\label{shreq}
\end{equation}
where $\cG\star$ is some radially symmetric convolution. My argument to
support this recommendation has three facets. Firstly, symmetry
considerations permit non-local operators in a model (\S2). Secondly,
non-local operators are naturally generated in systematic methods of
modelling nonlinear pattern evolution \cite{Roberts92b} (\S3). Indeed, both
Swift \& Hohenberg \cite{Swift77} and Bestehorn {\em et al}
\cite{Bestehorn89} naturally encountered non-local terms in their
derivations; but they heuristically argued to replace them by local
nonlinearities. Thirdly, the range of Fourier harmonics generated by the
nonlinearities is fundamentally different in two-dimensions than in
one-dimension (\S4). In the two-dimensions of a planform evolution problem
nonlinearities generate a continuous disc in Fourier space of harmonics,
whereasin
one-dimension the generated harmonics congregate in discrete lumps. This
difference requires a more sophisticated treatment of the two-dimensional
planform evolution problem, one that necessarily leads to non-local
nonlinearities.

\section{Symmetry arguments permit non-local effects}

The rationale behind the use of the Swift-Hohenberg equation~(\ref{sheq})
is firstly that for parameter values near critical, $\mu\approx 0$,
the spectrum of the linear terms, $\lambda= \mu- \left(k_0^2
-|\k|^2\right)^2$, matches, near the critical wavenumber $k_0$, the spectrum
of the physical problem under study. Secondly, that the cubic nonlinearity
is typical for the symmetry of many systems and is needed to stabilise
finite amplitude dynamics. The symmetry invoked is that of sign: $A\to -A$
leaves the Swift-Hohenberg equation~(\ref{sheq}) unchanged; as does $a\to
-a$ in the toy problem~(\ref{toyeq}), and $(\theta,w,z)\to (-\theta,-w,-z)$
in Rayleigh-Benard convection~(\ref{rbeq}). This symmetry is only
consistent with
odd functions. However, this symmetry applies to the field as a whole, not
to every part of it individually. Thus a wide variety of cubic functions
are potentially permissible: for example, non-local functions such as
$A(\x)^2A(\x+\vi)$ or $$\int K(\x,\x',\x'',\x''') A(\x')A(\x'')A(\x''')\,
d\x'd\x''d\x'''\,,$$ where $\x=(x,y)$ etc.

However, translational and rotational symmetry in space requires
that any non-local effects must be expressible as radially symmetric
convolutions. The local nature of typical models of physical dynamics
implies that harmonics are forced by locally expressible functions. But the
feedback from the harmonic to the critical modes, essential for
stabilisation of finite amplitude dynamics, need not be local. Indeed a
number of systematic studies have shown that memory, either temporal or
upstream, are needed in the low-dimensional modelling of forced dynamics
\cite{Cox91} or shear dispersion in varying channels
\cite{Smith83b,Mercer90}. In a planform evolution problem such a ``memory''
of horizontal structure, occurring through fluid convection for example,
would manifest itself as the non-local convolution in a cubic nonlinearity,
such as the $A\cG\star A^2$ term in~(\ref{shreq}) where $\cG$ is some
radially symmetric kernel.

\section{Non-local nonlinearities are natural}

The evolution of the Swift-Hohenberg equation contains a thin annulus of
critical modes, $|\k|\approx k_0$, near onset. It also contains a wide
variety of non-critical, exponentially damped modes, $|\k|\not\approx k_0$.
Thus the interesting long-term evolution of the critical modes is embedded
within the equation along with the evolution of many uninteresting modes.
It is for just such a scenario that I developed the concept of an embedded
centre manifold~\cite{Roberts92b}. There I show that adiabatic iteration,
namely the repeated application of adiabatic
elimination~\cite{Haken83,Vankampen85,Titi90}, effectively embeds the
critical modes of a slow manifold into the dynamics of a higher-dimensional
system. In a pattern evolution problem, the state space of the
higher-dimensional system consists of all the modes in the two-dimensional
plan,
whereas the slow manifold is composed of just the annular neighbourhood of
the critical modes.

For the toy problem~(\ref{toyeq}) with $\alpha_0=1-r$, $\alpha_1=2$ and
$\alpha_2=1$, adiabatic iteration leads to~(\ref{shreq}) as
a first non-trivial approximation where
\begin{equation}
	\left[(1-\nabla^2)^2-r\right]\cG=\delta(\x)\,.
	\label{toyg}
\end{equation}
For example, if $r=0$ then $\cG=\frac{1}{4\pi^2}K_0(\x)\star K_0(\x)$ in
two-dimensions and $\cG=\frac{1}{4}(1+|x|)\exp(-|x|)$ in one-dimension.

Now the left-hand side of~(\ref{toyg}) comes directly from the spectrum of
the exponentially decaying branches of the linearised problem. The only way
to avoid a non-local operator is if the solution of the analogue
of~(\ref{toyg}), in the given physical system, is a delta function,
$\cG\propto \delta(\x)$. Given the typically elliptic nature of dissipative
and spatially symmetric operators, this can only happen if the operator on
the left-hand side of the analogue to~(\ref{toyg}) is constant as a
function of wave-number $|\k|$ (or more
generally, constant on each branch). Generally this will not occur; for
example, in Rayleigh-Benard convection~(\ref{rbeq}) with stress-free
boundaries and $Pr=1$ the spectrum of the $m$th branch is
\begin{equation}
	\lambda=-m^2-|\k|^2\pm\sqrt{Ra}\frac{|\k|}{\sqrt{|\k|^2+m^2}}\,.
	\label{rbspec}
\end{equation}

Thus it is generic that systematic modelling naturally leads to non-local
nonlinearities such as shown in~(\ref{shreq}).

\section{Forced harmonics in Fourier space}

Here I argue that the richness of non-local nonlinearities are generally
necessary for an accurate model of a physical problem.  The argument
rests on the basic nature of the power spectrum in the two-dimensional
problem.
\begin{figure}
 \caption{schematic diagram in wave-number space of the leading order
nonlinear interactions between modes, $\k_1$ and $\k_2$, on the critical
circle $|\k|=k_0$ through the nonlinear generation of the harmonic at
point $B$.}
	\protect\label{wavenofig}
\end{figure}
As shown in Figure~\ref{wavenofig}, two critical modes with wavenumbers
$\k_1$ and $\k_2$ on the critical circle $|\k|=k_0$ interact through
quadratic nonlinearities to generate a harmonic at $B$, wavenumber
$\k_1+\k_2$. In the toy problem~(\ref{toyeq}) the relevant quadratic
nonlinearity is the $a^2$ term in the $b$ equation; in convection it is the
advection terms, $\q\cdot\nabla\q$ and $\q\cdot\nabla\theta$, on the
left-hand sides of~(\ref{rbeq}). Then such a forced harmonic interacts with
the $-\k_1$ critical mode, via the $ab$ term in~(\ref{toyeq}) or the
advection terms in~(\ref{rbeq}) for example, to generate a forcing of the
$\k_2$ component in the critical modes. By varying $\k_1$ and $\k_2$
independently around the critical circle, {\em every} wavenumber within the
disc $|\k|<2k_0$ is independently forced.\footnote{It is symmetry, for
example $a\to -a$, which ensures that two critical modes together do not
directly force another critical mode.}

Thus the amplitude spectrum of
typical planform evolution evolution is as shown in Figure~\ref{2dspecfig}.
\begin{figure}
	\caption{schematic typical amplitude spectrum of two-dimensional pattern
	evolution where the critical modes with $|\k|=k_0$ have typical amplitude
	$\epsilon$.}
	\protect\label{2dspecfig}
\end{figure}
Observe that this amplitude spectrum is qualitatively different from that
in one-dimension. In one-dimension the amplitude spectrum consists of
discrete lumps at integer multiples, $nk_0$, of the critical wavenumber.
For example, this property of the one-dimensional problem is crucial in
proofs that a Ginzburg-Landau equation is relevant to one-dimensional
pattern evolution\cite{Eckhaus93}. However, in two-dimensional problems the
amplitude spectrum is considerably richer. Consequently, in order to model
the leading order, physical interactions among all these modes, it is
{\em necessary} to determine the forced harmonics over the entire disk
$|\k|<2k_0$. Because these harmonics naturally decay with a rate which
depends upon wavenumber, then in physical space this determination can only
be done via a non-local convolution. For example, in the toy problem we
solve~(\ref{toyg}) for use in~(\ref{shreq}) in order to account for the
variations of the decay rate $-\lambda=-r+(1+|\k|^2)^2$ of the harmonics
$b$ over
the disc $|\k|<2k_0$.

The best that the Swift-Hohenberg equation~(\ref{sheq}) can do is to
approximate the functional dependence $\lambda(|\k|)$, as above or
in~(\ref{rbspec}), by a constant. It is easy to imagine problems where this
approximation would be inadequate. For example, if in the toy
problem~(\ref{toyeq}) we choose $\alpha_1=-4\alpha_2>0$, then the decay
rate of the harmonic $b$ has a minimum at wavenumbers $|\k|=\sqrt2 k_0$ and
so I expect that a square planform would be preferred because the harmonics
involved in the necessary interactions are not damped as strongly as other
stable modes. (In convection there can be a minimum in the decay-rate, but
for $Pr=1$, (\ref{rbspec}), it occurs for $|\k|<k_0$.) The Swift-Hohenberg
equation misses such subtleties, and as such it cannot be expected to be a
reliable model of any given physical problem. Instead, non-local terms based
on the wavenumber dependence of the decay of harmonics need to be used in
order to make reliable physical predictions from such a model.

Lastly, I give a further extremely formal argument for~(\ref{shreq}), one
based on the new notion of matching centre manifolds\cite{Watt94c} in order
to systematically develop models of specific problems.  The idea is to
develop a model, of lower-dimension than the original dynamical
system, whose slow manifold evolution matches that of the original system
to some order of analysis.  This principle is very like that of using
Pad\'e approximants to sum Taylor series.  Here I would propose the
modified Swift-Hohenberg equation~(\ref{shreq}) as a model, and then
determine the necessary constraints on the non-local convolution,
$\cG\star$, in order for the slow evolution of the critical modes
of~(\ref{shreq}) to be the same as those of the physical problem under
study.

Here the most restricted version of critical modes are those precisely on
the circle $|\k|=k_0$.  Thus we express the solution field as, for example,
\[ A(\x,t)=\int_0^{2\pi} Z_\phi(t)\exp(i\k_0(\phi)\cdot\x)d\phi
   +\cdots\,,
\]
for some complex amplitudes $Z$.  Concentrating only upon the nonlinear
interactions, the evolution of the slow modes would take the form
\[ \frac{\partial Z_\phi}{\partial t}
  =\cdots +\int_0^{2\pi} \beta(\phi,\psi) Z_\phi Z_\psi Z_{\pi+\psi}\, d\psi
\]
Recently, Edwards \& Fauve\cite{Edwards94b} have in essence used a
discretised version of of this, their equation~(8), in studying pattern
selection in the Faraday experiment. Indeed their interaction diagram,
Figure~4, is essentially the same as Figure~\ref{wavenofig} above. In order
for the matching to take place to cubic order, it is necessary that the
interaction coefficient, $\beta(\phi,\psi)$, for the physical problem and
for the model to be identical.  As explained before, this interaction
takes place through all the forced harmonics in the disc $|\k|<2k_0$. In
essence, $(\phi,\psi)$ appearing in the interaction integral above is a
straightforward parameterisation of this disc.  (For example, rotational
symmetry in the problems imply that $\beta$ is purely a function of the
difference
$\phi-\psi$.)  Thus, the only way for the
model~(\ref{shreq}) to accurately match a specific physical problem, is
for the interaction over all the harmonics in the disc to be accurately
represented.  Thus non-local nonlinearities are essential.

The only freedom this matching principle permits, up to third
order in the analysis, is the freedom to vary the Fourier transform of
$\cG$ for wavenumbers greater than $2k_0$. This cannot change the non-local
nature of the convolution.  In general, the Swift-Hohenberg equation
requires non-local modification.

\paragraph{Acknowledgements} I thank the Institut de Mecanique Statistic de
la Turbulence, Marseille, and the Institut Non Lin\'eaire de Nice for
their hospitality and support during the preparation of this work.

\end{document}